\documentstyle[11pt,newpasp,twoside,epsf]{article} 
\def\edcomment#1{\iffalse\marginpar{\raggedright\sl#1\/}\else\relax\fi} 
\reversemarginpar 
 
\begin{document} 
\title{Evolutionary properties of intermediate-mass stars} 
 \author{Santi Cassisi} 
\affil{INAF - Astronomical Observatory of Collurania, Via M. Maggini, 64100 Teramo, Italy} 
 
\begin{abstract} 
We briefly review the main problems related to the computation of the
evolution of intermediate-mass stars: the treatment 
of turbulent convection and the occurrence of blue loops during the core He-burning phase.  
It is shown that, in order to obtain more accurate and reliable
stellar models for this class of stars, one 
has to consider all possible theoretical and observational
constraints. This includes observations of low-mass stars to constrain
the treatment of envelope convection, and the analysis of the
pulsational behaviour of Cepheid stars.  
\end{abstract} 
 
\section{Introduction} 
 
In the last decade, thanks to the large advances in the computation of
opacity, equation of state (EOS) and nuclear reaction rates in stellar conditions, 
numerical stellar modeling  has been able to greatly improve our understanding of the star evolution throughout 
the H-R diagram. Strong confirmation of the reliability of evolution theory has come from the wonderful 
agreement between the predicted positions of stars in the Color-Magnitude Diagram (CMD) and observations. 
The pulsational behaviour of Cepheids and other radially oscillating stars has further confirmed the 
accuracy of stellar modeling for specific classes of stars. A detailed confirmation of some of the physics of 
stellar models has also been provided by the excellent agreement between the oscillation spectrum 
predicted by the standard solar model and helioseismological measurements. 
 
This notwithstanding, many aspects of stellar evolution theory have not been observationally 
confirmed, and thus deserve further investigations. In this respect, the evolutionary and structural 
properties of intermediate mass stars play a central role; as an
example, it is commonly believed that existing  uncertainties in the
theory of turbulent convection still greatly affect our understanding of the internal 
structure of these stars. 
 
The plan of this paper is as follows: in the next section we briefly discuss the reasons why 
stellar models for intermediate mass stars appear - in principle -
more reliable that models for low-mass stars, 
and outline the still-existing questions concerning their evolution. To this aim, 
we will consider observations of the pulsational properties of Cepheid stars. 
Specific attention will be devoted to highlight the \lq{sensitivity}\rq\ of the stellar 
model properties to the underlying physical assumptions. Conclusions follow. 
 
\section{The physics of intermediate-mass stars} 
 
The recent literature for intermediate-mass stars is quite rich (see Bono et al. 
2000 and references therein), and a detailed discussion of the main evolutionary and structural properties of these stars 
can be easily found elsewhere. Here we wish to shed light on the
questions one has to face when trying to model this class of stars. 
 
As stated in the previous section, the theoretical investigation of low-mass stars appears well supported by the 
results of helioseismology, but this is not the case for
intermediate-mass stars. The reason is not due to the difficulty 
in describing the microscopic physics at work in this kind of stars. In fact,
contrary to less massive objects, the evaluation of their thermodinamical properties is 
relatively simple, due to the fact that
intermediate mass stars are not affected, all along their major evolutionary phases, by electron degeneracy and 
other non-ideal effects such as Coulomb interaction 
The same applies to the computation of the radiative opacity, 
that is not affected by the significant uncertainties existing in the higher density and lower temperature regime. 
 
The evolutionary properties of intermediate-mass stars are strongly influenced by two open 
problems: 1) the extension of the convective core during the H-burning phase; 2) the - almost - unpredictable nature  
of the loop(s) in the CMD during the He central burning phase. These two
problems are closely connected and have large implications  
for the pulsational properties of intermediate mass stars when they cross the Cepheid instability strip. 
 
\section{The treatment of convection during the H-burning phase} 
 
Owing to the lack of a conclusive test for the adequacy of the current theory of convection, one 
can find a variety of different approaches to the inclusion of
convection in stellar models. The instability against 
turbulent convection is classically handled by means of the Schwarzschild 
criterion for a chemically homogeneous fluid. This well known
criterion is based on the comparison between the 
expected temperature gradient produced by the radiative transport of
energy, and the adiabatic one. It  is worth noticing 
that the proper estimate of the size of the convective region is then primarily dependent on the accuracy of the input 
physics. Any improvement of the adopted physics may produce a change in the estimated temperature gradient and, 
in turn modify the location of the boundaries of the convective regions. 
 
A second important point concerns the possibility that the motion of
the convective fluid elements is not drastically 
stopped at the stable region located just outside the formal
convective core. Although outside the Schwarzschild boundary a moving
fluid element is subject to a strong deceleration, it might be possible that a 
nonzero velocity is mantained along a certain length. This mechanical overshoot may induce a significant amount of  
mixing in the region stable against convection. 
Another important aspect which has to be taken into account is that a sizeable increase in the internal mixing and, hence 
a larger core could be also achieved as a consequence of rotationally
induced mixing (Meynet \& Maeder 2000). 
 
There are therefore at least three different reasons why the temperature gradient and, in turn the size of the 
convective cores can be really different that the one predicted by a \lq{canonical}\rq\ model\footnote{When 
referring to a canonical model, we mean a stellar structure computed by neglecting rotation, mechanical overshoot 
and accounting for the best physics {\sl currently} available.}. 
The question is: is the size of the convective core, as determined by the (classical) Schwarschild criterion 
able to properly match the observations, or it has to be \lq{artificially}\rq\ increased? 
 
This long standing problem has always been interpreted in terms of {\sl overshooting}, i.e. a proof of the 
existence or not existence of a specific phenomenon where the convective cells cross the classical border of the 
convective core\footnote{The extension of this non-canonical mixed region is usually defined in terms of a 
parameter $\lambda$ which sets the length - expressed as a fraction of
the local pressure scale height - which is crossed by the convective 
cells in the convectively stable region surrounding the convective core.}.  
However, when accounting for the various processes which, in principle, can produce the same 
effect, this interpretation appears meaningless. In our belief, it is
better to denote with  "overshooting" an unspecified mechanism(s) able
to increase the convective core size beyond the classical
Schwarzschild boundary. 
 
The case for a significant amount of overshooting has been presented many times both theoretically and 
observationally, but the result so far have been contradictory, even  
also those based on the analysis of stellar counts 
in young populous clusters (Testa et al. 1999; Barmina, Girardi \& Chiosi 2002; Brocato et al. 2003). The facts that are 
unarguably known concern only the basic theory: during the core H-burning phase the star has a larger He 
core, a brighter luminosity and a longer lifetime than in case of no overshooting. During the core He-burning 
phase, the luminosity is brighter, the lifetime is shorter and the blue loops are less extended than in the 
absence of overshooting. Theory therefore predicts that the mass-luminosity (M-L) 
relationship for stars crossing the Cepheid instability strip is largely affected by the amount of overshooting 
accounted for in the stellar evolution computations. The comparison of the theoretical M-L relationship with empirical 
data could, in principle, put tight constraints on the efficiency of this process. In the 
following, we will show that this is a feasible approach, but the results have to be cautiously treated. 
 
\subsection{The mass-luminosity relationship for Bump Cepheid stars} 
 
Keller \& Wood (2002, hereinafter KW) have used "bump" Cepheids in
order to probe the stellar M-L 
relationship for intermediate mass stars. By fitting 
empirical lightcurves for 20 bump Cepheids in the Large Magellanic Clouds with their own pulsational 
models, they estimated the intrinsic luminosity and mass of these variables. This allowed them to  
obtain a semi-empirical M-L relation. The comparison between their
data and theoretical 
results from the Padua group and from our own models is shown in figure 1. 
It is evident that semi-empirical data are systematically brighter than
canonical models, or alternatively, KW results support a mass $\approx15-20$\% 
lower than predicted by classical models, at a fixed luminosity. The
location of the data with respect 
to theoretical M-L 
relationships based on models accounting for overshooting seems to suggest the need for a degree of convective core 
overshooting with $\lambda_C\approx0.5$.  
 
\begin{figure} 
\plottwo{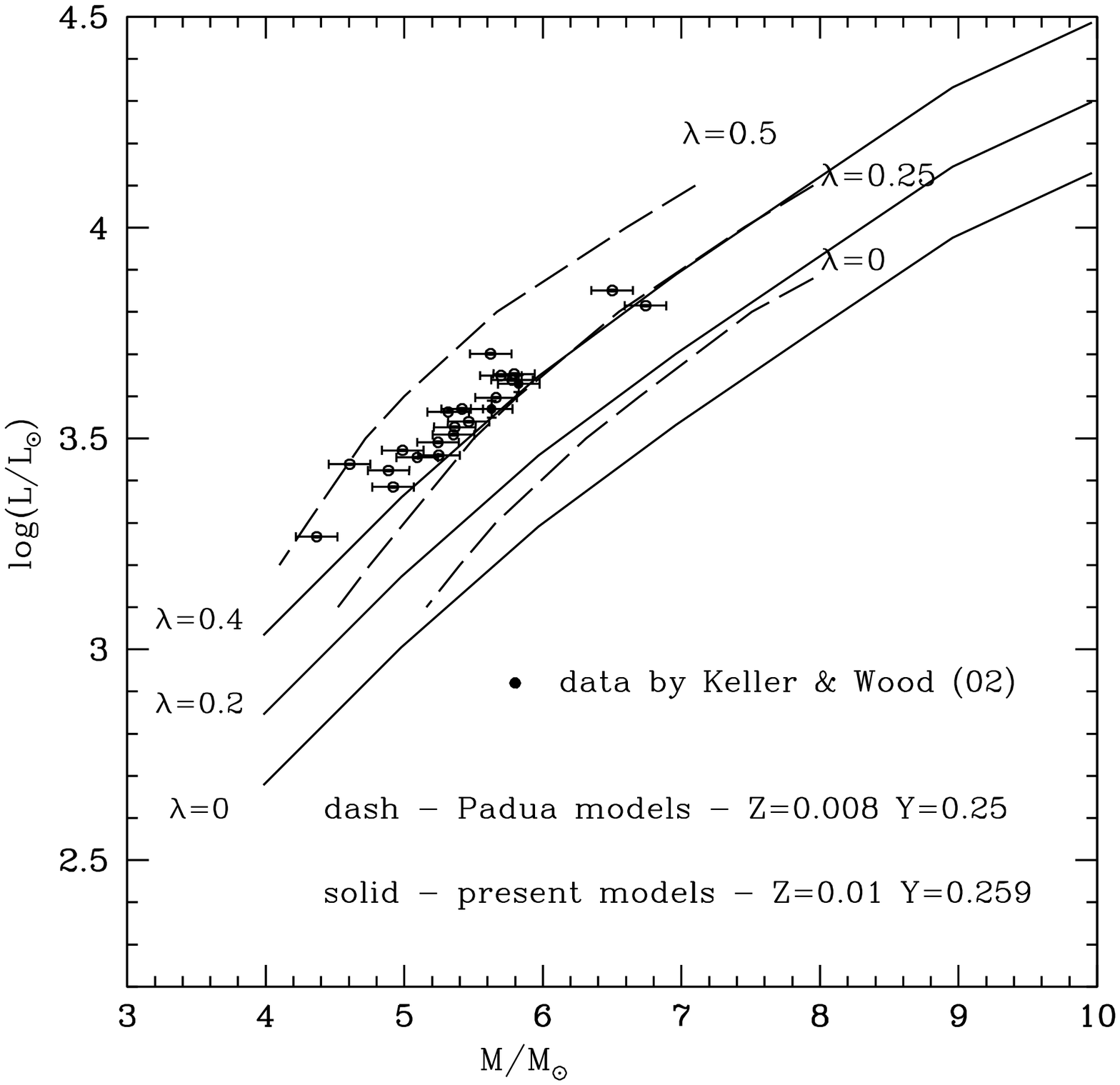}{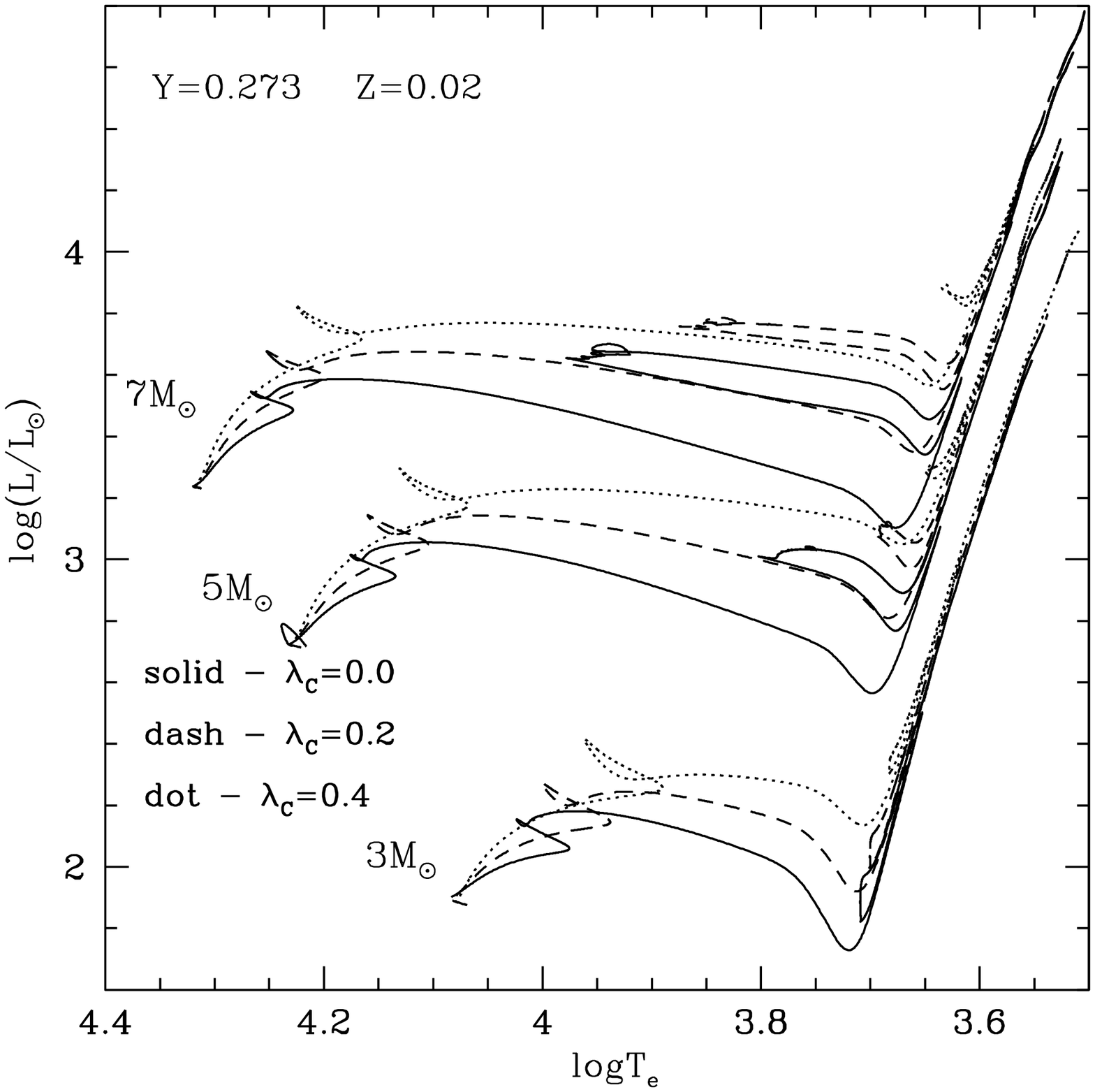} 
\caption{{\sl Left panel}: M-L relationship for the sample of bump Cepheids analized by KW. Overplotted are the 
theoretical M-L relationships provided by Padua models (see KW for details) and our own models for different (see labels) 
assumptions on the chemical composition and the core overshoot efficiency. {\sl Right panel}: H-R diagram for selected 
intermediate mass stars for different assumptions on the core convective overshoot efficiency.} 
\end{figure} 
 
Although, in our belief, the uncertainties in the M and L values as estimated by KW appear too small, their result  
seems to provide a plain support to the occurrence of a significant 
amount of core convective overshoot in intermediate-mass stars.  
However, this result has not to be uncritically accepted. In fact, one
has to take into account that by increasing the
overshoot efficiency to the value requested by KW results obtained by 
the blue loop excursion (see figure 1, right panel) is strongly reduced. 
One has therefore to check if models accounting for this overshoot efficiency are still able to 
cross the Cepheid instability strip. 
 
In figure 2 (left panel), we plot the minimum and maximum 
effective temperature reached during the blue loop by an intermediate
mass stellar model,
as a function of the core overshoot efficiency parameter $\lambda_C$;
we also display the $T_{eff}$ value at  
which the star spends more time during the blue loop, as a function of $\lambda_C$. 
The red and blue edges of the Cepheid 
instability strip are also shown. It is worth noticing that for a core overshoot efficiency larger than 
$\approx0.3$ the model does not cross the strip any longer. This suggests that 
it is not possible to claim, on the basis of the M-L relation alone, the need for a sizeable amount of core overshoot, without  
checking if that amount of core overshoot still allows the model to cross the instability strip. 
We will come back to this point in the next section; here we conclude by stating that if, as suggested by the KW 
analysis, a significant amount of core overshoot is really needed, one has to invoke some additional mechanism(s) 
in order to force the structure to cross the instability strip. 
 
Before closing this section, it is worthwhile to notice that a
suggested possible resolution of the discrepancy between 
evolutionary and pulsation mass for canonical models,
is to consider the occurrence of a quite efficient mass loss (Bono, Castellani \& Marconi 2002). 
However, this does not appear the right route for solving the problem at least for two reasons: 1) the requested  
amount of mass loss for solving the discrepancy is at odds with empirical estimates (Deasy 1988); 2) if 
the mass loss efficiency in these stars is really larger than currently expected, the effect on the blue loops  
of an high efficiency mass loss is to decrease both the $T_{eff}$ excursion and the mean luminosity. 
\begin{figure} 
\plottwo{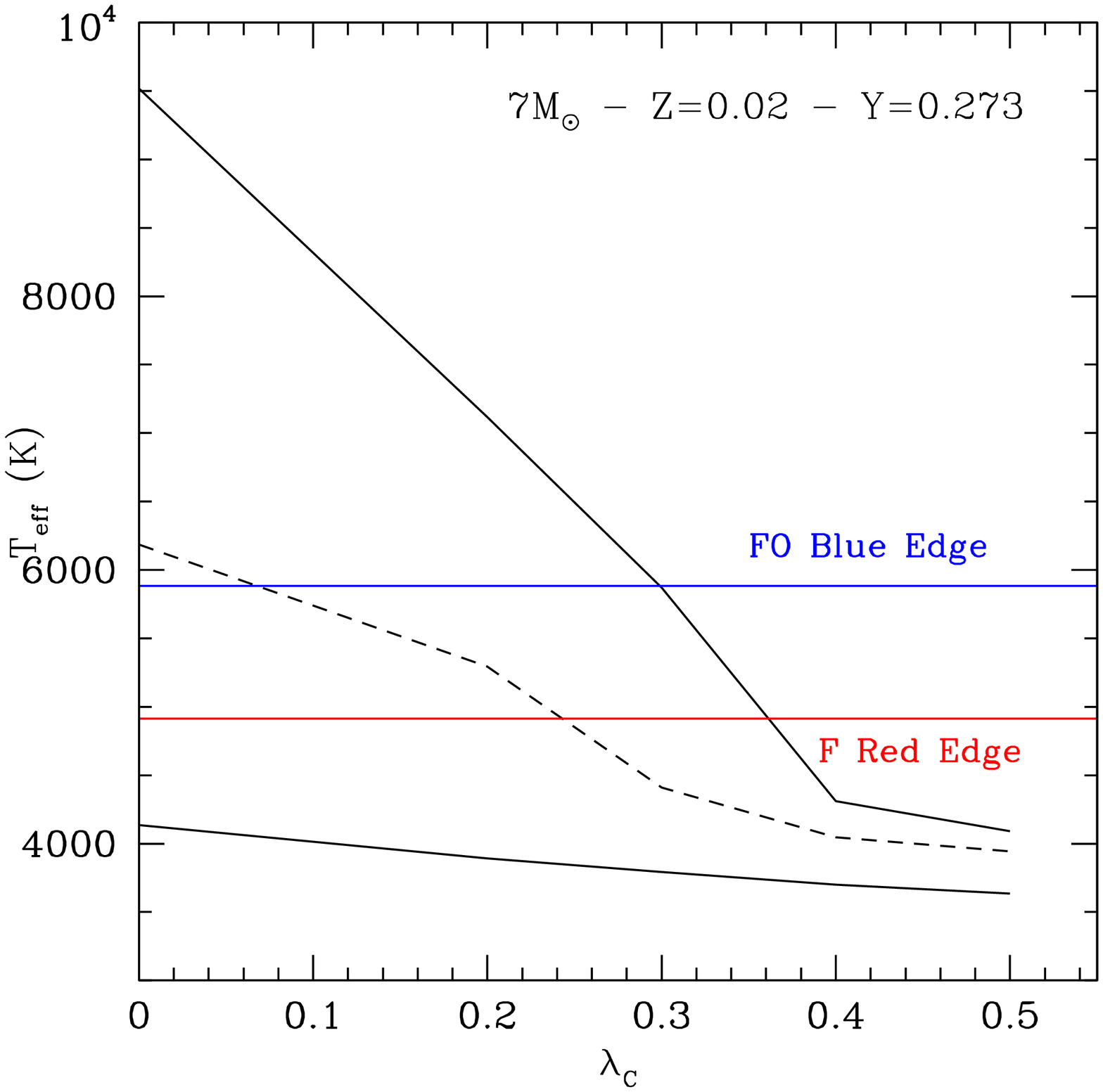}{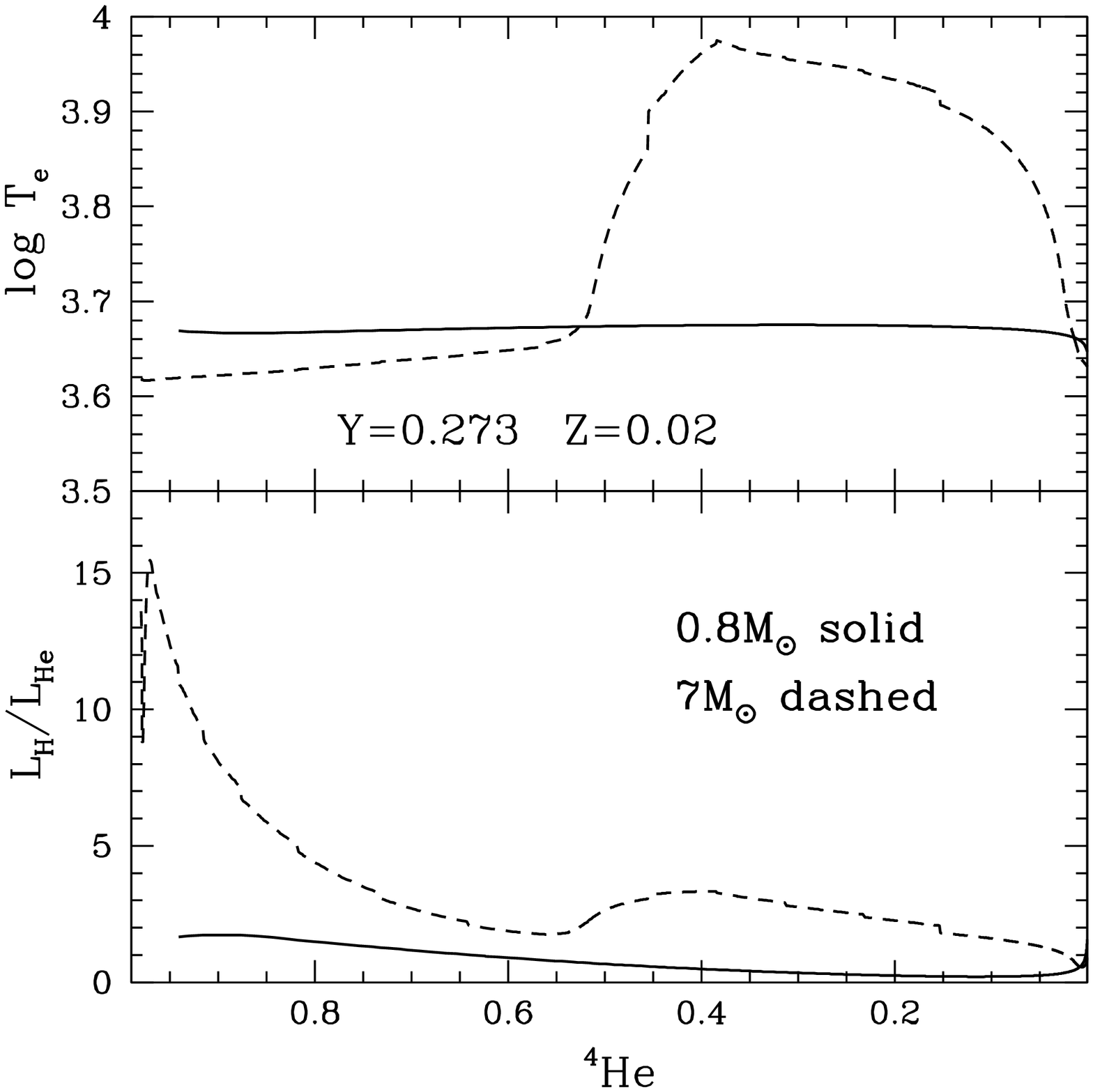} 
\caption{{\sl Left panel}: The run of the minimum, maximum and mean effective temperature during the 
blue loop of a 7$M_{\sun}$ model with solar composition as a function of the core overshooting efficiency. 
The location of the fundamental red edge and first overtone blue edge of the Cepheid instability strip is also 
shown. {\sl Right panel}: The behaviour of the effective temperature and of the ratio between the H-burning energy and 
He-burning one as a function of the central abundance of He in a $0.8M_{\sun}$ and $7M_{\sun}$ models.} 
\end{figure}

\section{"To loop or not to loop"} 
 
For a long time, the physical reasons for the existence of blue loops
challenged our understanding of stellar evolution. In this 
last decade, a big effort has been devoted to this subject; thanks
mainly to the pivotal work made by Renzini et al. (1992, hereinafter R92) and Renzini \& Ritossa (1994), we 
know better the physical reasons for the occurrence of blue 
loop(s) during the He burning phase. This notwithstanding, it is not yet possible to 
easily predict the behavior of an intermediate mass star model during this phase when 
changing some physical inputs and/or the physical assumptions adopted in the 
evolutionary computations. This is in contrast with what occurs for low-mass 
stars, for which the changes in their H-R location induced by any variation 
of the input physics can be easily predicted. The reason is that 
in intermediate mass stars the 
contribution to the stellar energy budget provided by the H-burning shell is 
significantly larger than in low-mass stars, and in addition the relative 
energy contribution of the He- and H-burning (see figure 2, right panel) changes significantly 
during the core He-burning phase. As a consequence, any variation  
of physical inputs able to modify the H-burning efficiency may either trigger or inhibit the loop. 
 
\begin{figure} 
\plottwo{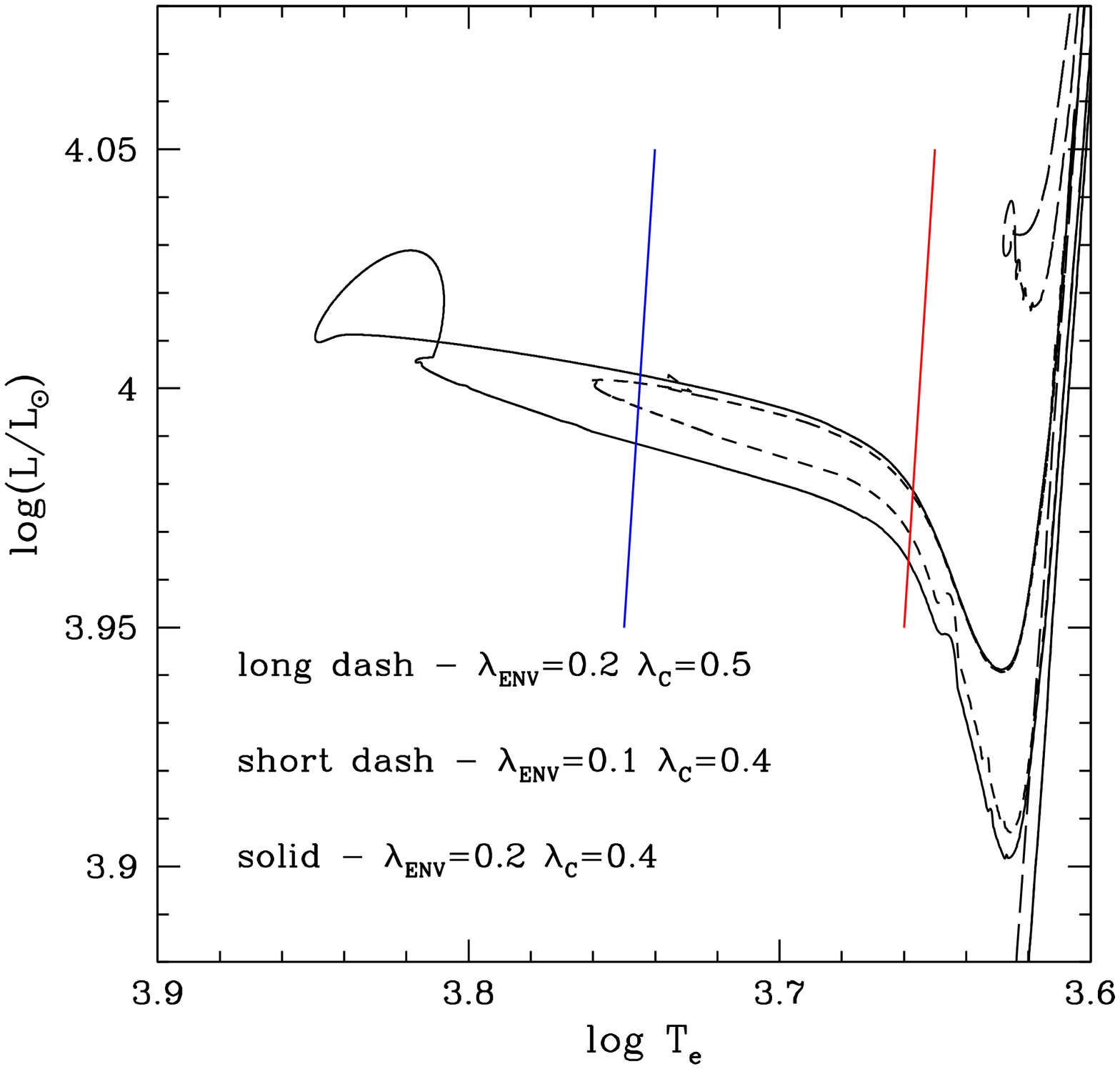}{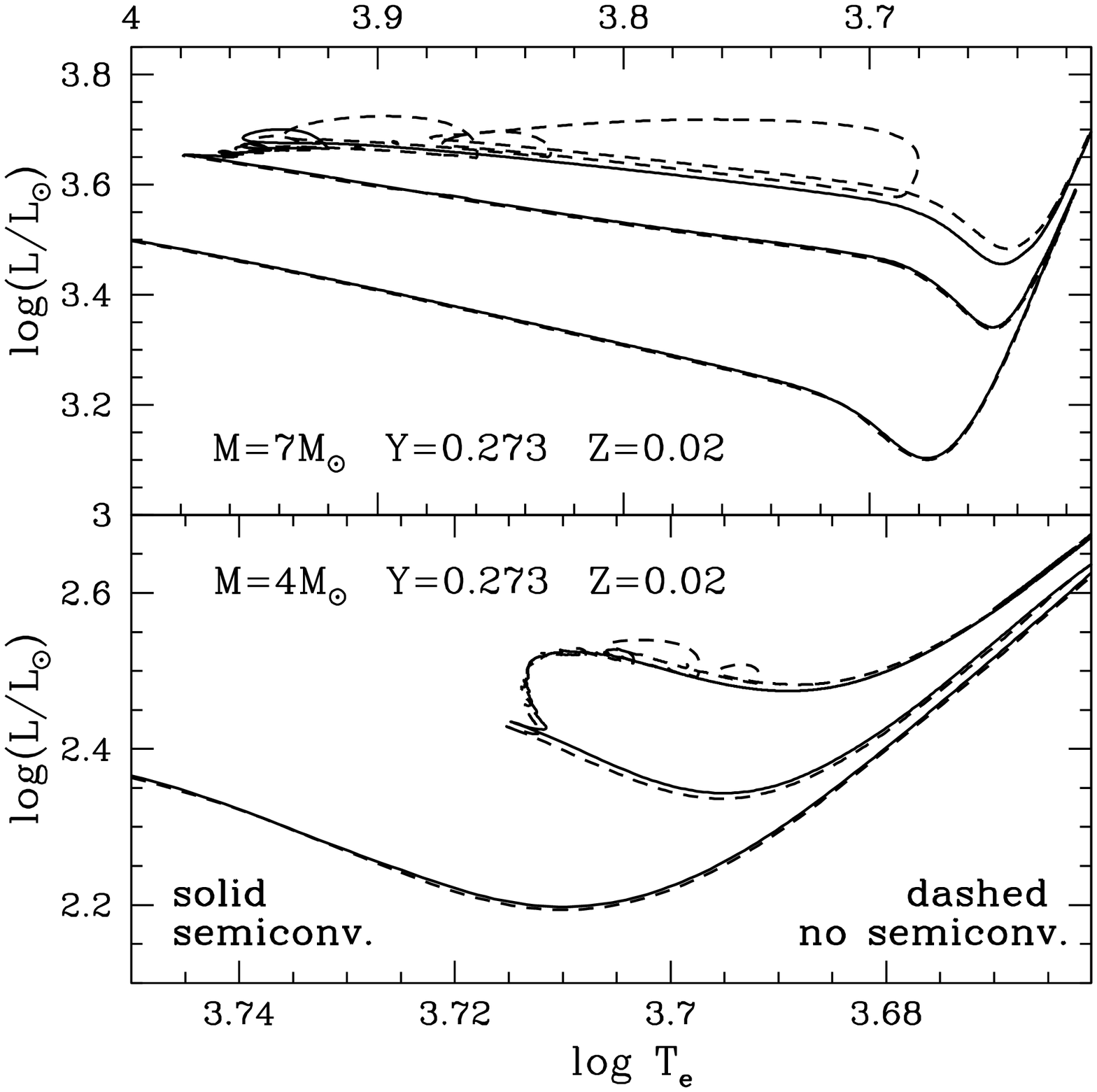} 
\caption{{\sl Left panel}: The location of the blue loop for a $7M_{\sun}$ model accounting 
for different efficiency of both core and envelope convective
overshooting, compared to the edges of the 
Cepheid instability strip. {\sl Right panel}: The H-R diagram location of the blue loops for two intermediate mass 
models computed alternatively by accounting for or neglecting the semiconvective region during the core He-burning 
phase.} 
\end{figure} 
 
An accurate analysis of the role of several factors known to affect the blue loops has been performed 
by R92 and we refer the interested reader to the quoted reference. In the following 
we address only the problem of the dependence of blue loops on convective envelope overshooting 
and semiconvection. 
 
\subsection{The consequence of convective envelope overshooting} 
 
The role played by convective envelope overshooting has been
investigated by several authors, e.g. Alongi et al.~(1991), Stothers \& Chin~(1991) 
and R92. Nevertheless, its effect still appears somewhat intricate. Evolutionary computations 
seem to suggest that a blue loop is apparently favoured by a sharper H profile in the envelope chemical stratification, 
as a consequence of the drastic change in the H-shell burning
efficiency (Cassisi \& Salaris 1997); this means that envelope 
overshooting, being able to produce this kind of sharper discontinuity during the first dredge up, may be able to trigger 
a loop that otherwise would not occur. However, as shown by R92, the general picture is more complicated, 
depending also on when the H-shell burning encounters this discontinuity - if this occurs after the central He ignition, 
envelope overshooting has no effect at all on the development of the blue loop. 
 
We have previously shown that the comparison between the semi-empirical M-L relationship for bump Cepheids and 
theoretical stellar evolution models does seem to indicate the need of accounting for a core overshooting of the 
order of $\lambda_C\approx0.5$; however, models accounting for this large overshoot efficiency are not able to cross  
the instability strip. If core overshooting really
occurs, it is reasonable to consider also the occurrence of envelope 
overshooting, whose efficiency promotes the occurrence of blue loops. In figure 3 (left panel) we  
show the shape of the blue loops for a selected intermediate mass model computed under several assumptions about both core 
and envelope overshooting, and their location with respect to the boundaries of the instability strip. 
For a core overshooting efficiency of the order of ${\lambda}_C=0.4$, an envelope overshooting 
efficiency of the order of $\lambda_{ENV}=0.1$ is enough to push the model inside the instability strip.  
Nevertheless, for ${\lambda}_C=0.5$ (see previous section) theoretical evidences show that an envelope overshooting  
efficiency $\lambda_{ENV}>0.2$ is requested. 
 
We can now ask if observational data can help constraining the efficiency of envelope overshooting. The 
answer is positive, and this can be done accounting for the brightness of the Red Giant luminosity function bump (see Salaris, 
Cassisi \& Weiss 2002) in galactic globular clusters: the comparison between observational 
data and evolutionary models clearly 
suggests that {\sl if} envelope overshooting is really useful for improving the match between theory and observations, then 
its efficiency has to be lower than $\lambda_{ENV}=0.2$. It is therefore evident that the amount of envelope 
overshooting needed by intermediate mass stars is in marginal agreement - if not ruled out - by evolutionary evidences for 
low-mass Red Giant stars. Of course, one has to bear in mind that the efficiency of envelope overshooting - if any - can 
depend on the stellar mass, and in principle the amount of overshooting in intermediate mass stars could be different than 
that in low-mass stars. However, in our belief, the constraint provided by low-mass stars has to 
be properly accounted for. If this is the case, one has to invoke some different physical mechanism(s) able to favour 
the occurrence of extended blue loops in intermediate mass models accounting for huge amounts of core overshooting. 
 
\subsection{The role of semiconvection} 
 
On the basis of preliminary numerical experiments R92 suggested that the presence of a semiconvective region  
at the edge of the He-burning convective core, by changing the rate at which He is mixed into the core,  
could have significant effects on the occurrence of the blue loops. 
More specifically, the larger is the semiconvective region, the lower 
is the chance for the occurrence of blue loops.  
In order to explore this scenario, we computed selected models for
intermediate mass stars by alternatively accounting for or neglecting
semiconvection. 
In figure 3 (right panel) we show the comparison between these models. One can easily notice that, at variance 
with previous indications, the effect of semiconvection on the morphology of the blue loops is quite negligible. This 
occurrence being due to the fact that the semiconvective region in
these stars is only a small fraction of the whole convective 
core ($\approx10$\% for a $4M_{\sun}$ model and $\approx1$\% for a $7M_{\sun}$ one). 
 
\section{Conclusions} 
 
Even if the analysis of the evolutionary and structural properties of intermediate-mass stars has been the subject of many 
investigations during last years, there remain many aspects of their evolution which have to be still fully 
understood. 
In this respect, the study of the efficiency of turbulent convection at the \lq{canonical}\rq\ boundaries of both 
convective core and envelope has to be reanalized by properly accounting also for the constraints coming from low-mass 
stars in globular clusters, as well as asteroseismological measurements (Guenther 2002), whose impact will increase in the near 
future thanks to various planned space missions. As far as this last point
is concerned, it is worthwhile remembering that 
the accurate analysis of pulsational spectra of field white dwarfs 
(Metcalfe, Salaris \& Winget 2002) is starting to provide a powerful tool for 
investigating the efficiency of mixing in the He core of their intermediate-mass progenitor. 
One can also predict an improved understanding of the evolution of intermediate-mass 
stars when both observations of pulsational properties and theory will be analized together. 
 
\acknowledgements 
I warmly thank M. Salaris for the interesting discussions on these topics as well as for an accurate reading 
of the manuscript. I wish also to thank A. Pietrinferni for all the help provided. It is also a pleasure to
thank G. Bono for his valuable suggestions and constant encouragement.

\end{document}